# Maximum-Entropy-Rate Selection of Features for Classifying Changes in Knee and Ankle Dynamics During Running

Garry A. Einicke, *Senior Member, IEEE,* Haider A. Sabti, David V. Thiel, *Senior Member, IEEE,* and Marta Fernandez

*Abstract*—This paper investigates deteriorations in knee and ankle dynamics during running. Changes in lower limb accelerations are analyzed by a wearable musculo-skeletal monitoring system. The system employs a machine learning technique to classify joint stiffness. A maximum-entropy-rate method is developed to select the most relevant features. Experimental results demonstrate that distance travelled and energy expended can be estimated from observed changes in knee and ankle motions during 5 km runs.

## I. INTRODUCTION

MUSCLE fatigue is commonly defined as a reduced ability to maintain force or power output during prolonged physical activity. Disturbances in the concentrations of muscle lactate, hydrogen, potassium and calcium ions are linked with fatigue [1]. Fatigue is also associated with a reduction in the metabolic uptake of oxygen by muscle cells during prolonged exercise. Fatigue monitoring is important for three reasons. In fatigue management, sportspeople may need to be rested and revitalized to maximize performance. Second, to aid injury prevention as tendon and muscle functions become impaired with overuse. Apart from subjective feelings of soreness and tiredness, continuing sporting activities during fatigue can adversely affect lower limb abilities to absorb shock and to maintain joint stability. Compromised joint stability can in turn increase the likelihood of strains, sprains, stress fractures and falls. Third, to diagnose post-injury rehabilitation. Objective indicators of joint stability can provide information about the efficacy of support measures (*e.g.*, taping, strengthening, conditioning) and for monitoring recovery.

For example, a video motion analysis of competitive runners during a 5 km road race [2] found that deterioration in running technique (*i.e.*, reductions in step length and cadence) can occur from 2.4 km. A biomechanical analysis of fatigue-related foot injury mechanisms appears in [3] with an observed fatigue of the peroneus longus, pre-tibial and triceps surae muscles attributed to a deterioration in foot stability and increased vulnerability to ankle sprains. Magnetic resonance imaging has been used to monitor the healing of a runner that had suffered a sacral fatigue-type fracture [4].

Musculo-skeleto monitoring systems have been developed for squat and leg extension exercises [5] - [9], cycling [8] - [11], walking [3] and running [2], [12] - [24]. Fatigue patterns were characterized in [5] as the drop of instantaneous median frequency of electromyographic (EMG) signals for vastus lateralis, rectus femoris, vastus medialis and biceps femoris muscles during squats. This lent support to existing theories of muscle compensatory functions in patients with anterior cruciate ligament (ACL) deficiency. The instantaneous mean frequency of surface EMG signals for the same muscle group has been used to diagnose fatiguing contractions [6]. The mean frequency of EMG amplitude spectra for dumbbell biceps curls, dumbbell lateral rises and squats was analyzed in [7]. It reported a decreased average mean frequency as muscles becomes weaker due to the reduced conduction velocity of muscle fibres. Fatigue can be estimated by observing kinematic changes in squat exercises from optical motion capture data and correlated with individuals' subjective assessments [8]. Changes in EMG median frequencies for cyclists' biceps femoris and gastrocnemius muscles have also been discussed in [8], [11]. The authors of [9] studied squat kinematics using a ten-camera motion analysis system. They showed that exhausting exercise produced kinematic changes at the trunk and pelvis. A running modeling study [13] suggested that muscle fatigue should not significantly change ground reaction force peaks which may increase the level of soft-tissue vibrations. An assessment of inertial measurement units for monitoring runners is detailed in [14]. Treadmill runners exhibited lowered step frequency as they fatigued, whereas outdoor track runners decreased their movement efficiency. A study into the impact of running velocity during treadmill running [15] suggested that horizontal and vertical force production could diagnose an athlete's strengths and weaknesses. Volunteers equipped with eight joint reflective markers were filmed running on a treadmill until exhaustion in [16]. It was observed that ankle range of motions, maximal knee flexion during stance, maximal knee flexion

G. A. Einicke is with Griffith University and CSIRO, Australia (phone: +61 7 3327 4615; fax: +61 7 3327 4566; e-mail: g.einicke@ Griffith.edu.au). H. A. Sabti is with Griffith University (email: haider.sabti@gmail.com). D. V. Thiel is with Griffith University (email: d.thiel@griffith.edu.au). M. Fernandez is with Griffith University, Australia and the University of the Basque Country, Spain (email: martafernandez010@gmail.com).

during swing, and knee range of motion all increased at the end the run. Plantarflexion, ankle inversion and eversion were observed to increase in a comparable running experiment [17]. It is mentioned in [18] that fatigue may cause muscle imbalances and alteration of joint stability. The paper [18] also reported increased angular velocities of knee flexion and ankle flexion after prolonged running. A study of runners [19] found that increased average hip excursion and knee flexion occur during fatigue. Tests with runners on a treadmill [20] concluded that decreases in muscle strength of knee extensors and flexors occur before and after foot impact. The foot motion of runners equipped with shoe-based inertial measurement units is analyzed in [21]. It emerged that the range of foot motion in the sagittal plane increased in the final phases of 10-km-long running races. The authors of [22] used a six-camera motion analysis system to test runners. They recommended balanced training to reduce peak valgus and internal rotation moments to lower anterior cruciate ligament injury risk. It is similarly noted in [23] - [24] that a gradual increase in knee flexion at heel strike occurs during running.

The developments described herein seek to assist with detecting the onset of fatigue. A wearable musculo-skeleto monitoring system is described that differs from the approaches in [2] - [24]. The developed system includes a number of sensor modules which can be worn within pockets of Velcro straps or sports garments. The sensor modules have a triaxial accelerometer (MMA7260QT) having a sensitivity of 2 g, a 3.6 V, 240 mAh battery that is sufficient to support several hours of data logging and Wifi communications. The sensors were programmed to store accelerometer measurements during running and transfer the data to an external computer using a CP2102 UART interface, see [25], [26]. A photo of the sensor module is shown in Fig. 1.

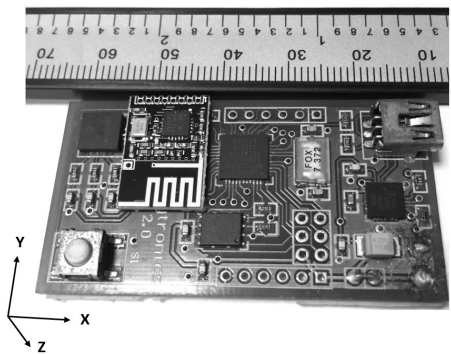

Fig. 1. Photo of the sensor module.

It is desired to calculate bio-mechanical features from the sensor data for classifying the onset of fatigue. However, two ambiguities require to be overcome:

(1) The features reported in [2] – [24] have different statistical significance; and

(2) The comparative evaluation of features within [27] – [30] can involve a high calculation cost.

The variation in feature significance within [2] – [24] is probably due to individuals having subtly different running styles, depending on anatomical variations, training and injury histories. Some runners have close to symmetrical lower limbs and normal arches of the foot, whereas others may have tendencies to over pronate or supinate. While sprinters seek to maximize power output, endurance athletes are more concerned with maintaining an efficient running style.

Since irrelevant features degrade classification accuracy and increase calculation cost, a large number of feature selection techniques have been developed, see the surveys [27] – [30]. Measures for comparing feature subsets include Pearson correlation, mutual entropy, cross entropy, information gain and Kullback-Leibler divergence [27] – [30]. The computation cost of evaluating such measures for each $2^n - 1$ combination of $n$ features can be prohibitive.

The above ambiguities are resolved by assembling a vector of candidate features and employing a machine learning technique to select the best-performing ones. Calculating the first three sample moments (variance, skewness and kurtosis) of knee and ankle triaxial accelerometer data results in $n = 18$ candidate features. A maximum-entropy-rate feature selection procedure is developed which departs from those reviewed in [27] – [30]. To minimize calculation cost, the features are ranked individually and the maximum average per-feature entropy over $n - 1$ subsets is found. The selected features are employed within a weighted-minimum-distance classifier. This enables a runner's on-line measurements to be classified against previously-recorded data for estimating their distance travelled and energy expended.

The developments described herein are confined to estimating a recreational runner's level of fatigue. An open problem that remains is predicting the risks of injuries corresponding to observed levels of fatigue. It may be possible in the future to extend the developed method to include higher-order sample moments, additional sensors, professional athletes, different age groups, and other activities such as walking, cycling and swimming.

The remainder of this paper is organized as follows. The data collection method and estimation of runners' energy expenditure are described in Sections IIA and IIB, respectively. Second IIC discusses the use of statistical moments for monitoring variations in knee and ankle motions. An optimal filter for removing noise in moment measurements is set out in Section IID. Section IIE describes the main contribution of the paper, namely, a maximum-entropy-rate feature selection procedure. This procedure can identify the subset of features whose associated classification probabilities are approximately equal. A maximum-entropy-rate feature subset and probability distribution vector subset are employed in a weighted-minimum-distance classifier which is discussed in Section IIF. Some experimental results are presented in Section III. It is demonstrated that a runner's distance travelled and energy expenditure can be estimated with 15% average classification error. The conclusion follows in Section IV.

## II. DATA COLLECTION AND ANALYSIS METHOD

### A. Data Collection Method

Three of the authors volunteered to run laps around a grassy oval whilst wearing accelerometer sensors during a summer at Griffith University. The runners' gender, age, weight, height and body mass index (BMI) attributes are summarized in Table I.

The running took place during a summer in which the temperature varied from 23 to 36° C and the humidity ranged from 65 to 75%. To minimise the risk of dehydration and heat stress, the runners stopped for several seconds after each lap for a drink of water. This work was conducted under a code for the responsible conduct of research (Ethics Approval ENG/20/13/HREC).

Triaxial accelerometer measurements were recorded using the wireless sensors described in [25], [26]. Two accelerometer sensors, denoted by 1 and 2, were worn on the lateral sides of the right knee and right ankle, respectively. Axes 1, 2 and 3 denote accelerations in the X (mediolateral or left-right), Y (superior-inferior or up-down), and Z (anterior-posterior or forward-backward) directions, respectively. The three-axis accelerometer data were sampled at 100 Hz.

TABLE I
RUNNERS' PHYSICAL ATTRIBUTES

| Runner | Gender | Age | Weight | Height | BMI |
|---|---|---|---|---|---|
| 1 | male | 59 y | 90 kg | 1.8 m | 29 |
| 2 | female | 27 y | 52 kg | 1.63 m | 20 |
| 3 | male | 31 y | 77 kg | 1.74 m | 25 |

### B. Energy Expenditure

The Griffith University oval perimeter was measured to be 0.455 km using a GPS-Glonass receiver. The researchers ran 11 laps, *i.e.*, 5 km. The duration of the 5-km-long runs were typically about 25 minutes. It is well known that a low-intensity warm-up can boost muscle coordination and cardiovascular efficiency. However, the runners did not perform any warm-up exercises (in order to minimise time under the sun).

The measurements were segmented into $N$ subintervals. The accumulated kinetic energy expenditure after the $k^{th}$ subinterval is estimated as $\sum_{i=1}^{k} 0.5 m v_i^2$, in which $m$ is the runner's weight (from Table 1) and $v_i$ is their observed average speed over subinterval $i$. The kinetic energies expended by runners 1 – 3 are plotted in Fig. 2. A fatigue index can be calculated from expended energies. For example, from the difference between maximal and minimal power output, and from the rate of decline in power output [31]. As shown in the figure, our runners did not decrease their speed and so a fatigue index can instead be defined as the fraction of accumulated kinetic energy expenditure during a run, *i.e.*,

$$\left( \sum_{i=1}^{k} 0.5 m v_i^2 \right) \left( \sum_{i=1}^{N} 0.5 m v_i^2 \right)^{-1} \times 100\% . \quad (1)$$

The machine learning system developed in the following involves a training step and a classification step. In the training step, an individual is required to run two or more times, during which features are calculated from accelerometer data and recorded. In the classification step, the individual runs again, and their on-line accelerometer features are compared with previous data to estimate their (time-varying) subinterval. Their estimated subinterval can be cross-referenced against their energy expenditure (from Fig. 2) or a fatigue index such as (1).

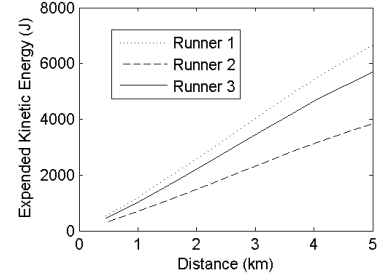

Fig. 2. Accumulated kinetic energy expenditure by Runners 1 - 3.

### C. Statistical Moments

Let $E\{Z_k^{(i,j)}\}$ denote the mean of the set of measurements $Z_k^{(i,j)}$ from accelerometer $i$ along axis $j$ within subinterval $k$, in which $E\{.\}$ is the expectation operator. Three statistical moments, namely, the sample variance $(\sigma_k^{(i,j)})^2 = E\{(Z_k^{(i,j)} - E\{Z_k^{(i,j)}\})^2\}$, sample skewness $\alpha_k^{(i,j)} = E\{((Z_k^{(i,j)}) - E\{Z_k^{(i,j)}\})/\sigma_k^{(i,j)})^3\}$ and sample kurtosis $\beta_k^{(i,j)} = E\{((Z_k^{(i,j)}) - E\{Z_k^{(i,j)}\})/\sigma_k^{(i,j)})^4\}$ were calculated for each subinterval.

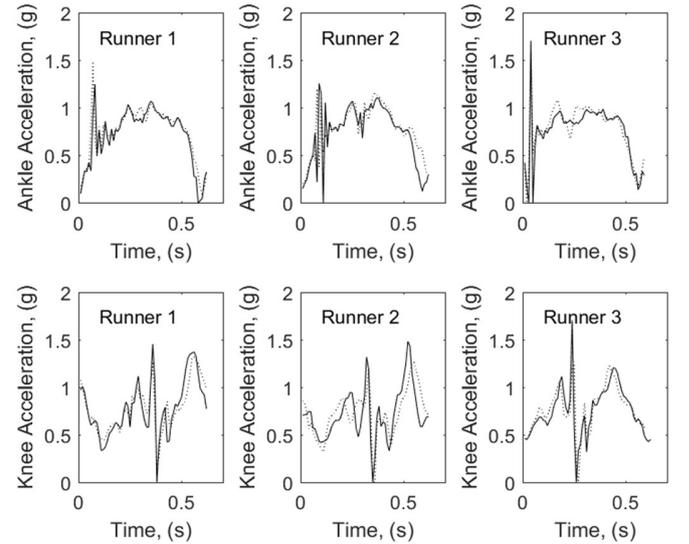

Fig. 3. Example ankle and knee acceleration sensor outputs along the superior-inferior or up-down direction during a gait cycle for Runners 1, 2 and 3. The dotted and solid lines correspond to Laps 1 and 11, respectively.

It was observed for the above-mentioned runners that several moment trajectories exhibited decreasing or increasing trends with energy expended during the run. For example, for Runner 1, it was observed that: (a) ankle acceleration variances along axes 1 - 3 increased; and (b) knee acceleration variances along axes 2 - 3 increased. Observation (a) suggests that increased movement occurs in ankles during running-induced fatigue. This is attributed to fatigued muscles and tendons in the foot losing their ability to cushion shock. Some example superior-inferior ankle and knee accelerations during Lap 1 and Lap 11 are shown in Fig. 3. The ankle acceleration peaks at mid-stance correspond to active muscle forces. The peaks in the first quarter of the gait cycle correspond to the absorption of ground contact force, which are seen to be larger during the final lap. Observation (b) is attributed to increased knee flexion and transfer of impact shock increasing during fatigue – as illustrated by the knee acceleration peaks at mid-stance.

### D. Filtering of Moments

As mentioned above, biomechanical features having different statistical significance are reported in [2] – [24]. Therefore, the above three sample moments were calculated for all six accelerometer signals, which yielded an 18-element candidate feature vector

$$z_k = [(\sigma_k^{(1,1)})^2, (\sigma_k^{(1,2)})^2, (\sigma_k^{(1,3)})^2,$$
$$\alpha_k^{(1,1)}, \alpha_k^{(1,2)}, \alpha_k^{(1,3)},$$
$$\beta_k^{(1,1)}, \beta_k^{(1,2)}, \beta_k^{(1,3)},$$
$$(\sigma_k^{(2,1)})^2, (\sigma_k^{(2,2)})^2, (\sigma_k^{(2,3)})^2,$$
$$\alpha_k^{(2,1)}, \alpha_k^{(2,2)}, \alpha_k^{(2,3)},$$
$$\beta_k^{(2,1)}, \beta_k^{(2,2)}, \beta_k^{(2,3)}]^T. \quad (2)$$

The initial ordering of the components within (2) is arbitrary. This feature vector will subsequently be re-ordered and truncated (as described in Section IIIE).

An individual's current state of fitness may be estimated by classifying their current feature vector against the mean of previously recorded feature vectors. However, during each lap the runners travelled over slightly different trajectories and undulations, which resulted in noisy feature vector observations. The presence of noise can adversely affect classifier performance. To remove noise, linear regression was used to fit a least-squares line to previously recorded feature measurements, whereas on-line feature observations were filtered as follows.

Optimal minimum-error-variance filtered estimates, $x_k$, were obtained from the first-order, one-step recursion

$$x_k = (I - L)Ax_{k-1} + Lz_k, \quad (3)$$

where $L \in \mathbb{R}$ is a filter gain and $A \in \mathbb{R}$ is a prediction coefficient, see p. 83 of [32]. An unbiased estimate of the prediction coefficient is given by $A = E\{z_{k+1}z_k^T\}(E\{z_kz_k^T\} - R)^{-1}$, where $R$ is the measurement noise variance. The filter gain is calculated as $L = P(P + R)^{-1}$, where $P \in \mathbb{R}$ is the solution of the Riccati equation $P = A(P - LP)A + Q$, in which an unbiased input variance, $Q$, is estimated from $Q = E\{z_kz_k^T\} - R - A(E\{z_kz_k^T\} - R)A$.

The accelerometer measurements were normalized and segmented into $N$ sub-intervals. The first-order filter (3) was applied to the on-line observations of (2), in which $R$ was obtained from the moments' sample variances. As mentioned above, feature trajectories and statistical significance vary from one individual to another. Therefore, a method for selecting salient features was developed below.

### E. Maximum-Entropy-Rate Feature Selection

The selection of relevant features remains an open problem, see the surveys [27] – [30] and the references therein. A formal definition of feature relevance appears in [27]. Eliminating irrelevant features can lead to improved classification accuracy and reduces the classification calculation cost. As many of the feature selection techniques in [27] – [30] have high computational overheads costs, a low-calculation-cost method is developed in the following.

Suppose that the means of two previously recorded feature vector sets $u_k = [u_{1,k},...,u_{n,k}]^T$ and $v_k = [v_{1,k},...,v_{n,k}]^T$ of the form (2) are available for subinterval classes $k \in [1, N]$. A discrepancy vector $d = [d_1,...,d_n]^T$ is constructed for $j \in [1, n]$ feature components, where

$$d_j = \frac{D([u_{j,1},...,u_{j,N}],[v_{j,1},...,v_{j,N}])}{\sum_{k=1}^{n} D([u_{k,1},...,u_{k,N}],[v_{k,1},...,v_{k,N}])} \in \mathbb{R}, \quad (4)$$

in which $D(.)$ denotes a distance measure such as the Euclidean norm or the Mahalanobis distance. The discrepancies (4) represent the probability that $[u_{j,1},...,u_{j,N}]$ is furthest from $[v_{j,1},...,v_{j,N}]$, given feature component $j$. Let $\bar{d} = [\bar{d}_1,...,\bar{d}_n]^T$, $\bar{d}_j = \dfrac{1 - d_j}{\sum_{k=1}^{n}(1 - d_k)}$, denote a vector of probabilities that $[u_{j,1},...,u_{j,N}]$ is nearest to $[v_{j,1},...,v_{j,N}]$, given feature components $j \in [1, n]$. High values of $\bar{d}_j$ suggest that feature component $j$ is more relevant for discriminating between classes. Conversely, low values of $\bar{d}_j$ suggest that feature component $j$ is less relevant for class discrimination. Let $p = [p_1,...,p_n]^T$ denote a vector whose components are those of $\bar{d}$ into sorted into non-increasing order, i.e.,

$$p_1 = \max(\bar{d}), \quad p_i = \max_{i \in [2,n]}\{p_i \in \bar{d} \mid p_i \geq p_{i-1}\}. \quad (5)$$

The feature selection task amounts to finding a length $L \leq n$ subset of the ordered probability distribution vector $p$ that satisfies a performance measure. The Shannon entropy, which is calculated as $-\sum_{i=1}^{L} p_i \log p_i$, refers the amount of uncertainty that is present. A suitable performance measure is the maximum entropy rate, *i.e.*, the maximum of the average (per feature) entropy

$$H_L = -\frac{1}{L}\sum_{i=1}^{L} p_i \log p_i. \qquad (6)$$

An iterative procedure for finding the most relevant length-$L$ subset of $p$ that maximizes (6) is set out below.

*Procedure 1:* Assume that the components of the distribution vector $p$ have been sorted into non-increasing order, *i.e.*, (5).

Step 1: Set $L = n$.
Step 2: Evaluate (6).
Step 3: If $L = 2$ or

$$\frac{1}{L-1}\sum_{i=1}^{L-1} p_i \log p_i < p_L \log p_L \qquad (7)$$

stop the iterations. Otherwise, set $L = L - 1$, normalize the probability distribution vector subset, *i.e.*, $p = \frac{[\bar{d}_1,...,\bar{d}_L]^T}{\sum_{i=1}^{L}\bar{d}_i}$ and go to Step 2.

It can be seen that the left hand side of the inequality (7) is the average entropy for $L - 1$ feature components. Thus, if condition (7) is false, the $L^{th}$ feature component is deemed to be irrelevant and eliminated. The procedure requires the entropy rate (6) to be only evaluated over $2 \leq L \leq n$ since $H_0 = 0$. A maximum entropy rate of $H_L = \frac{1}{L}\log L$ is attained in cases where the first $L$ elements of $p$ are equal. Thus, the procedure can identify the subset of features whose associated classification probabilities are approximately equal.

*F. Weighted-Minimum-Distance-Classification*

Suppose the following.

(i) An individual ran along the same route two or more times and their feature measurements (2) have been recorded.

(ii) The feature measurements have been normalized to unity variance (so that they have similar scales).

(iii) Linear regression has been used to fit a line to previously recorded feature measurements (to remove noise).

(iv) Procedure 1 has been used to select most relevant subset of features.

(v) The same person runs on another occasion and the filter (3) processed their relevant subset of normalized feature measurements (to remove noise).

Consider the problem of identifying the subinterval index $k$ of a new feature measurement $x_k$. The unknown $k$ can be estimated using the weighted-minimum-distance classifier

$$k = \underset{k\in[1,N]}{\arg\min}\ D(p \otimes x_k, p \otimes \bar{x}_k), \qquad (8)$$

where $\bar{x}_k$ is the fitted previously recorded feature, $D(.)$ is a distance measure and $\otimes$ denotes an elementwise multiplication. The inclusion of $p$ within (8) can yield a performance benefit because it pays greater weight to the more relevant features. In this application it has been found that improved estimates of $k$ can be obtained by classifying consecutive features $[x_{k-\ell}, ..., x_k] \in \mathbb{R}^{L\times\ell}$, with measurement lag $\ell \geq 1$, weighted by $p$, *i.e.*,

$$k = \underset{k\in[1,N]}{\arg\min} D(p\otimes[x_{k-\ell},...,x_k], p\otimes[\bar{x}_{k-\ell},...,\bar{x}_k]), \qquad (9)$$

where $[\bar{x}_{k-\ell}, ..., \bar{x}_k] \in \mathbb{R}^{L\times\ell}$ is the mean of previously recorded features and. Using $\ell$ measurements within (9) can improve average classification performance when consecutive features and their means are approximately equidistant.

It has been found here that Procedure 1 selected three feature components, in which case (8) corresponds to the minimum distance between three-dimensional vectors, and (9) corresponds the minimum distance between three-dimensional vectors averaged over $\ell$ measurement lags. The average time to perform the filtering (3), sorting (5), maximum entropy selection (Procedure 1) and lag-$\ell$ weighted-minimum-distance classification (9) was approximately 0.03 s on a 9.1 GHz quad-core processor. This suggests that a mobile device equipped with a 2.3 GHz quad-core processor would be able to perform these on-line calculations in about 0.1 s.

III. RESULTS AND DISCUSSION

The $n = 18$ moments were calculated over $N = 44$ quarter-lap or 113.6-m sub-intervals for each individual's 5-km run. The above procedure was applied to the discrepancies between the raw measurements (2) and the corresponding least-squares line fits. This procedure employed the maximum entropy rate criterion to select the three most-relevant features for each runner, which are listed in Table II. Some example feature trajectories for Runners 1 – 3 are shown in Fig. 4. The figure demonstrates that some features exhibit increasing or decreasing trends, which are exploited using the machine learning technique described above.

Changes in knee and ankle dynamics with fatigue have been previously studied in [17] - [19], [23] - [24]. For example, runners can present increased ankle eversion during fatigue [23] which could be caused by weakening of the tibialis anterior or tibialis posterior muscles. A transfer of foot eversion to tibial rotation could cause changes in knee moments. It is noted in [17] - [19], [23] - [24] that a gradual increase in knee flexion at heel strike occurs during running fatigue, which contributes to an increased tibial impact acceleration. Planta flexor muscle fatigue results in a decline in the ankle range of motion [24]. The combination of increased ankle eversion [23], increased tibial impact shock and planta flexor fatigue [19], [24] could result in changed ankle moments.

TABLE II
SELECTED FEATURES & RMS ERRORS

| Runner | Selected features | Lag 0 | Lag 1 | Lag 2 | Lag 3 | Lag 4 |
|---|---|---|---|---|---|---|
| 1 | $(\sigma_k^{(2,2)})^2$ $\alpha_k^{(2,2)}$ $\beta_k^{(2,2)}$ | 11% | 13% | 11% | 12% | 10% |
| 2 | $(\sigma_k^{(1,3)})^2$ $\beta_k^{(1,2)}$ $\beta_k^{(2,2)}$ | 26% | 25% | 24% | 24% | 23% |
| 3 | $(\sigma_k^{(1,1)})^2$ $(\sigma_k^{(2,1)})^2$ $(\sigma_k^{(2,3)})^2$ | 8% | 8% | 9% | 10% | 10% |
| Mean | | 15% | 15% | 15% | 15% | 15% |

A Euclidean norm distance measure was used in the weighted-minimum-distance classifier (9) to estimate each person's subinterval index during a second run. The estimated subinterval index can then be cross-referenced against distance travelled, energy expended (from Fig. 2) and a fatigue index (such as (1)). The resulting subinterval index estimation root-mean-square (RMS) errors are listed in Table II. It can be seen that the RMS errors are lower for Runners 1 and 3. That is, the onset of fatigue appears to be more distinct for runners having greater body mass. This may be due heavier runners being exposed to higher ground reaction forces that increase the loading of knee and ankle joints, which over time result in reduced shock absorbing capacity [12]. This interpretation is supported by Fig. 4, which shows that feature trajectory trends are slightly more distinct Runner 1 than for Runner 2. It can be seen from the table that it is possible to infer a runner's distance travelled and energy expended with an average error of 15%.

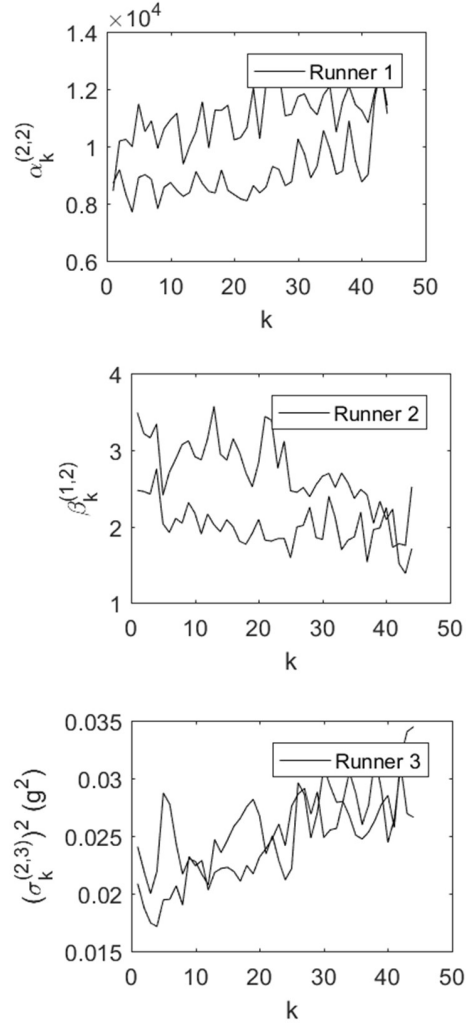

Fig. 4. Example selected feature trajectories versus subinterval index $k$ for Runner 1 (top), Runner 2 (middle) and Runner 3 (bottom).

The methods described in Section II have a low calculation cost - which suits fatigue monitoring in the field. The developments can probably be refined in the following ways.

(a) The Euclidean norm distance measure does not exploit class distribution knowledge. Using means of covariances from multiple feature measurements within distance measures (e.g., Mahalanobis distance) may yield improved classifier performance.

(b) The above feature selection method relies on features being ranked individually. Ranking pairs or triples of features may improve feature selection.

(c) The trade-off between average classification accuracy and the number of subinterval classes could be optimized. An average accuracy benefit arises when there are fewer classes to discriminate between, and the sample moments better approximating the actual moments with increasing number of samples.

Nevertheless, it has been observed that differences tend to occur with the placement and slippage of sensors, and the paths followed from one run to another. Therefore, some estimation performance variations are expected to remain.

## IV. Conclusion

Prolonged sporting activities can lead to tendon/muscle impairment and fatigue. Continuing exercise during fatigue can adversely affect lower limbs' abilities to absorb shocks and maintain joint stability which increases the likelihood of injuries. This motivates monitoring changes in knee and ankle looseness or stiffness to detect the onset of fatigue.

A musculo-skeleto monitoring system has been developed which uses machine learning to classify changes in knee and ankle accelerations. Three statistical moments, namely, variance, skewness and kurtosis of three-axis accelerations serve as features. Since irrelevant features can degrade classification accuracy, a low-calculation-cost method has been developed to select an individual's most relevant features. This method involves ranking the features individually and finding the maximum entropy rate subset. The selected features appear to be consistent with prior knowledge about changes in ankle eversion, knee flexion angle and tibial impact forces with fatigue. A weighted-minimum-distance classifier was then employed to discern changes in feature trajectories.

The developed system was used by three volunteers having considerably different physical attributes who ran 5-km around an oval on two occasions. It was found that a weighted-minimum-distance classifier was able to estimate a runner's distance travelled and energy expended with an average error of 15%.

## References

[1] G. C. Bogdanis, "Effects of physical activity and inactivity on muscle fatigue", *Frontiers in Physiology*, vol. 3, no. 142, 15 pp., May 2012.

[2] B. Hanley and L. Smith, "Effects of fatigue on technique during 5 km road running", *Proc. ISBS Conference Archive*, vol. 1, no. 1, pp. 382 – 385, 2009.

[3] A. Gefen, "Biomechanical analysis of fatigue-related foot injuring mechanisms in athletes and recruits during intensive marching", *Medical & Biological Engineering & Computing.*, vol. 40, pp. 302 – 310, Mar. 2002.

[4] K. Knobloch, L. Schreibmueller, M Jagodzinski, J. Zeichen and C. Krettek, "Rapid rehabilitation programme following sacral stress fracture in a long-distance running female athlete", *Archives of Orthopaedic and trauma Surgery.*, vol. 127, iss. 9, pp. 809 – 813, Nov. 2007.

[5] P. Bonato, M.-S. S. Cheng, J. Gonzalez-Cueto, A. Leardini, J. O'Connor and S. H. Roy, "EMG-based measures of fatigue during a repetitive squat exercise", *IEEE Eng. Med. Biol. Mag.*, vol. 20, no. 6, pp. 133 – 143, Nov./Dec. 2001.

[6] F. Molinari, M. Knaflitz, P. Bonato and M. V. Actis, "Electrical manifestations of muscle fatigue during concentric and eccentric isokinetic knee flexion-extension movements", *IEEE Trans. Biomed. Eng.*, vol. 53, no. 7, pp. 1309 – 1316, Jul. 2006.

[7] G. Biagetti, P. Crippa, A. Curzi, S. Orcioni and C. Turchetti, "Analysis of the EMG signal during cyclic movements using multicomponent AM–FM decomposition", *IEEE J. Biomed. and Health Informatics*, vol. 19, no. 5, pp. 1672 – 1681, Sep. 2015.

[8] M. Karg, G. Venture, J. Hoey and D. Kulić, "Human movement analysis as a measure for fatigue: A hidden Markov-based approach", *IEEE Trans. Neural Systems and Rehab. Eng.*, vol. 22, no. 3, pp. 17 – 23, May. 2014.

[9] B. K. Weeks, C. P. Carty and S. A. Horan, "Effect of sex and fatigue on single squat kinematics in healthy young adults", *BMC Musculoskeletal Disorders*, vol. 16, no. 271, Sep. 2016.

[10] M. Knaflitz and F. Molinari, "Assessment of muscle fatigue during biking", *IEEE Trans. Neural Systems and rehab. Eng.*, vol. 11, no. 1, pp. 17 – 23, Mar. 2003.

[11] J. B. Dingwell, J. E. Joubert, F. Diefenthaeler, and J. D. Trinity, "Changes in muscle activity and kinematics of highly trained cyclists during fatigue", *IEEE Trans. Biomed. Eng.*, vol. 55, no. 11, pp. 2666 – 2674, Nov. 2008.

[12] S. A. Bus, "Ground reaction forces and kinematics in distance running in older-aged men", *Medicine & Science in Sports & Exercise*, vol. 35, no. 7, pp. 1167 – 1175, 2003.

[13] A. A. Nikooyan and A. A. Zadpoor, "Effects of muscle fatigue on the ground reaction force and soft-tissue vibrations during running: A model study", *IEEE Trans. Biomed. Eng.*, vol. 59, no. 3, pp. 797 – 804, Mar. 2012.

[14] C. Strohrmann, H. Harms, C. Kappeler-Setz, and G. Troster, "Monitoring kinematic changes with fatigue in running using body-worn sensors", *IEEE Trans. Information Tech. in Biomed.*, vol. 16, no. 5, pp. 983 – 990, Sep. 2012.

[15] M. Brughelli, J. Cronin and A. Chaouachi, "Effects of running velocity on running kinetics and kinematics", *J. Strength and Conditioning Research*, vol. 25, no. 4, pp. 933 – 939, 2011.

[16] L. De Lucca and S. I. L. Melo, "Relationship between running kinematic changes and time limit at vVO$_{2max}$", *Brazilian Journal of Kinanthropometry and Human Performance*, vol.14, no.4, 2012.

[17] J. Ryu, "The effect of fatigue caused by running time on the kinematic parameters of the lower uncertainty", *Proc. 8th Biennial Conf. Canadian Society of Biomechanics*, pp. 316 - 317, 2003.

[18] E. Kellis and C. Liassou, "The effect of selective muscle fatigue on sagittal lower Limb kinematics and muscle activity during level running", *J. of Orthopaedic & Sports Physical Therapy*, vol. 39, no. 3, pp. 210 – 220. Mar. 2009.

[19] J. Mizrahi, O. Verbitsky, E. Isakov, D. Daily, "Effect of fatigue on leg kinematics and impact acceleration in long distance running", *Human Movement Science*, vol. 19, pp. 139 – 151, 2000.

[20] E. Kellis, A. Zafeiridis, I. G. Arimidis, "Muscle coactivation before and after the impact phase of running following isokinetic fatigue", *J. of Athletic Training*, vol. 46, no. 1, pp. 11-19, Jan 2011.

[21] E. Medina, N. Palomares, A. Page, B. Bazuelo-Ruiz, "Analysis of kinematic patterns in runners. An approach based on inertial sensors and functional data analysis. *Proc. 33$^{rd}$ Int. Conf. Biomechanics in Sports*, Poitiers, France, June 2015.

[22] J. L. Cochrane, D. G. Lloyd, T. F. Besier, B. C. Elliot, T. L. Doyle and T. R. Ackland, "Training affects knee kinematics and kinetics in cutting maneuvers in sport", *Medicine and Science in Sports and Exercise*", vol. 42, no. 8, pp. 1535 – 1544, Aug. 2010.

[23] I. F. Koblbauer, K. S. van Schooten, E. A. Verhagen and J. H. van Dieen, "Kinematic changes during running-induced fatigue and relations with core endurance in novice runners", *J. of Science and Medicine in Sport*, vol. 17, pp. 419 – 424, 2014.

[24] M. Giandolini, P. Gimenenz, J. Temsi, P. J. Arnal, V. martin, T. Rupp, J.-B. Morin, P. Samozino, G. Y. Millet, "Effect of the fatigue induced by a 110-km ultramarathon on tibial impact acceleration and lower leg kinematics", *PLOS One*, no. 0151687, Mar. 2016.

[25] A. J. Wixted, D. V. Thiel, A. G. Hahn, C. J. Goore, D. B. Pyne and D. A. James, "Measurement of energy expenditure in Elite athletes using MEMS-based triaxial accelerometers", *IEEE Sensors J.*, vol. 7, no. 4, pp. 481 - 488, Apr. 2007.

[26] H. A. Sabti and D. V. Thiel, "Self-calibrating body sensor network based on periodic human movements", *IEEE Sensors J*, vol. 15, pp. 1552 - 1558, 2015.

[27] A. Blum and P. Langley, "Selection of relevant features and examples in machine learning", *Artificial Intelligence*, pp. 245 - 271, 1997.

[28] P. Kumbhar and M. Mali, "A survey on feature selection techniques and classification algorithms for efficient text classification", *Int. J. Science and Research*, vol. 5, iss. 5, pp. 1267 – 1275, May 2016.

[29] K. Mani and P. Kalpana, "A review on filter based feature selection", *Int. J. of Innovative Research in Computer and Communications Engineering*, vol. 4, iss. 5, pp. 9146 – 9156, May 2016.

[30] L. Wang, Y. Wang and Q. Chang, "Feature selection methods for big data informatics: A survey from the search perspective", *Methods*, vol. 111, pp. 21 – 31, 2016.

[31] M. N. Naharudin, A. Yusof, "Fatigue index and fatigue rate during an anaerobic performance under hypohydrations", *PLOS One*, vol. 8, iss. 10, 7 pp., Oct. 2013.

[32] G. A. Einicke, *Smoothing, filtering and prediction: Estimating the past, present and future*, InTech, Rijeka, Croatia, Feb. 2012.